\documentclass[aps,prd,onecolumn,preprintnumbers,groupedaddress,showpacs,
nofootinbib,amssymb]{revtex4}
\usepackage{graphicx}
\usepackage{epsfig}
\usepackage{bm,,color}
\usepackage{amssymb}
\usepackage{float}
\usepackage{amsmath}
\usepackage{cancel}

\allowdisplaybreaks[4]
\newcommand{\e}{\mathrm{e}}
\newcommand{\nn}{\nonumber \\}

\begin{document}

\title{The dark universe future and singularities:  the account of thermal and
quantum
effects}

\author{Shin'ichi~Nojiri$^{1,2}$, 
Sergei~D.~Odintsov$^{3,4,5}$
}

\affiliation{ $^{1)}$ Department of Physics, Nagoya University,
Nagoya 464-8602, Japan \\
$^{2)}$ Kobayashi-Maskawa Institute for the Origin of Particles
and the Universe, Nagoya University, Nagoya 464-8602, Japan \\
$^{3)}$ ICREA, Passeig Luis Companys, 23, 08010 Barcelona, Spain \\
$^{4)}$ Institute of Space Sciences (IEEC-CSIC) C. Can Magrans
s/n, 08193 Barcelona, Spain \\
$^{5}$ Inst. Physics, Kazan Federal University, Kazan 420008, Russia}

\begin{abstract}

The knowledge of the universe future is of fundamental importance for
any advanced civilization.
We study the future of the singular dark universe where thermal effects
due to the Hawking radiation on the apparent horizon of the FRW universe 
are taken in the consideration.
It is shown that the dark universe which ends up as finite-time Type I and
Type III singularity or the infinite-time Little Rip singularity transits to 
the finite-time Type II singularity thanks to account of the thermal effects.
However, Type II and IV singular universe does not  change
the qualitative behavior.
The combined account of the quantum and thermal effects shows that depending
on the specific features of the universe only one of the effects is dominant.
When (conformal matter) quantum effects are dominant, the future singularity 
is usually removed while for dominant thermal effects the universe final state is 
Type II singularity.

\end{abstract}
\maketitle

\section{Introduction}

Theoretical discovery of accelerating dark energy universe significally
changed our knowledge about the universe future. It is rather well-known
that the dark energy epoch may be qualitatively understood as  universe
filled by the exotic effective fluid with negative pressure.
Depending on its structure and in correspondence with observable bounds,
current dark universe may show phantom ($w_\mathrm{eff} < -1$),
de Sitter ($w_\mathrm{eff}=-1$) or quintessential ($-1<w_\mathrm{eff}<-1/3$) 
behaviour where $w_\mathrm{eff}$
is the effective universe EoS parameter. So far the precise understanding in
which dark era we live is still lacking.
However, for  big sub-class of phantom or quintessential universes it
turns out  that  the universe ends up in some sort of future singularity.
It is fundamentally important for any advanced civilization to know what 
happens with the universe in the future even this event  is rather distant 
(few dozen billion years).

The most known type of finite-time future singularity is related with phantom
evolution  and is  called Big Rip singularity. In this case the universe ends
up   very rapid expansion and any extended object will be destroyed by the tidal
force some million years before reaching the Rip time \cite{Caldwell:1999ew}.
Apart from this most strong singularity there are few soft singularities which
are classified as Type II, III, and IV singularities.
For the Big Rip singularity, the Hubble rate $H$ diverges in a finite time and 
the time $t_\mathrm{Rip}$ that the divergence occurs is called the Big Rip 
time. 
Note that this is the classical consideration. However, taking into account
different related effects (like quantum effects, strong electromagnetic
fields, condensation, etc.) may qualitatively change the classical
consideration and give the realistic picture of the future universe. 
First of all, let us remind
that the large Hubble rate $H$ means the large temperature of the universe.
The Hawking radiation effectively should be generated at the apparent
horizon of the FRW universe \cite{Gibbons:1977mu,Cai:2008gw}.
Eventually, it should give the important contribution
to the energy-density of late-time universe, especially right before the
Rip time. In other words, at large temperature which may even diverge at
the Rip time, there should appear  thermal
radiation. Recently in Ref.~\cite{Ruggiero:2020piq}, it was argued that
the kind of cyclic cosmology might be realized instead of the Big Rip 
singularity due to effect of thermal radiation.
The purpose of this paper is to study  what could really happen with
the future singular dark universe when the effect of  thermal radiation 
is included.
In the next section we consider the dark universe with Type I, II, III, and IV
finite-time future singularities as well as the infinite-time Little Rip
singularity in the presence of thermal effects due to the Hawking radiation. We
demonstrate that for Type II and IV singular universe there is no qualitative
effect to its final state due to the thermal radiation.  
Type I and III singular universes as well as the Little Rip universe ends 
up as Type II singularity due to thermal effects. 
The third section is devoted to the account of both, quantum and
thermal effects to future singularities. We show that quantum effects as a rule
remove future singularity creating the  non-singular universe. 
When quantum and thermal effects are taken into consideration, 
depending on the specific features of the theory (particles content, fluids, 
temperature, etc.) only one of the effects becomes dominant.
For instance, when the thermal effects are dominant, the universe ends up 
at Type II singularity in the same way as without quantum effects. 
Finally, the summary and some outlook are given in the last section.

\section{Finite-time future singularities in the dark universe: the account of
thermal effects}

We start from  a spatially-flat FRW universe,
\begin{equation}
\label{JGRG14}
ds^2 = - dt^2 + a(t)^2 \sum_{i=1,2,3} \left(dx^i\right)^2\, .
\end{equation}
Here $a(t)$ is a scale factor.
We consider dark energy epoch when the effective
equation of state (EoS) of the universe is around $-1$.
In this case, the accelerating universe may evolve in one of the following
ways: phantom evolution, quintessential evolution, and de Sitter expansion
($w_\mathrm{eff}=-1$). What happens with such universe in the future?
In principle, depending on specific aspects of time-dependent effective EoS of
the universe any future is possible including deccelerating or ever-expanding
universe. Let us consider here the sub-class of dark energy universes which
lead to finite-time future singularities.

Let us remind that the FRW equations for general relativity (GR)
coupled with general perfect fluid
with the pressure $p$ and the energy-density $\rho$ are given by
\begin{equation}
\label{JGRG11}
\rho=\frac{3}{\kappa^2}H^2 \, ,\quad
p= - \frac{1}{\kappa^2}\left(3H^2 + 2\dot H\right)\, .
\end{equation}
Here $H\equiv \dot a/a$.

For the dark energy universes ending at finite-time singularity it has been
developed the classification of such singularities in Ref.~\cite{Nojiri:2005sx}
(see also \cite{Odintsov:2018uaw}):
\begin{itemize}
\item Type I (``Big Rip'') Ref.~\cite{Caldwell:1999ew}:
This is a characteristic crushing type singularity, for which
as $t \to t_s$, the scale factor $a(t)$, the total effective pressure
$p_\mathrm{eff}$ and the total effective energy density $\rho_\mathrm{eff}$
diverge strongly, that is, $a \to \infty$, $\rho_\mathrm{eff} \to \infty$,
and $\left|p_\mathrm{eff}\right| \to \infty$.
For  works on this type of singularity, see
Refs.~\cite{Caldwell:2003vq,Nojiri:2003vn,Elizalde:2004mq,Faraoni:2001tq,
Singh:2003vx, Wu:2004ex,Sami:2003xv,Stefancic:2003rc,Chimento:2003qy,
Zhang:2005eg,Dabrowski:2006dd,Nojiri:2009pf,Capozziello:2009hc,
BeltranJimenez:2016dfc}.
\item Type II (``sudden''): This type of singularity is milder
than the Big Rip scenario, and it is also known as a pressure singularity,
firstly studied in Refs.~\cite{Barrow:2004xh,Nojiri:2004ip},
and later developed in \cite{Barrow:2004he,FernandezJambrina:2004yy,
BouhmadiLopez:2006fu,Barrow:2009df,BouhmadiLopez:2009jk,Barrow:2011ub,
Bouhmadi-Lopez:2013tua,Bouhmadi-Lopez:2013nma,
Chimento:2015gum,Cataldo:2017nck},
see also \cite{Balcerzak:2012ae,Marosek:2018huv}.
Here, only the total effective pressure diverges as $t \to t_s$,
and the total effective energy density and the scale factor remain finite,
that is, $a \to a_s$, $\rho_\mathrm{eff} \to \rho_s$ and
$\left|p_\mathrm{eff}\right| \to \infty$.
\item Type III : In this type of singularity, both the total effective pressure
and the total effective energy density diverge as $t \to t_s$, but the scale
factor remains finite, that is, $a \to a_s$, $\rho_\mathrm{eff} \to \infty$ 
and $\left|p_\mathrm{eff}\right| \to \infty$.
Then Type III singularity is milder than Type I (Big Rip) but stronger than 
Type II (sudden).
\item Type IV : This type of singularity is the mildest from a phenomenological
point of view. It was discovered in Ref.~\cite{Nojiri:2004pf} and further
investigated in
\cite{Nojiri:2005sx,Nojiri:2005sr,Barrow:2015ora,Nojiri:2015fra,
Nojiri:2015fia,Odintsov:2015zza,Oikonomou:2015qha,
Kleidis:2017ftt}.
In this case, all the aforementioned physical quantities remain finite as
$t \to t_s$, that is, $a \to a_s$, $\rho_\mathrm{eff} \to 0$,
$\left|p_\mathrm{eff}\right| \to 0$,
but higher derivatives of the Hubble rate, $H^{(n)}$ $\left( n\geq 2 \right)$
diverge.
This singularity may be related with
the inflationary era, since the Universe may smoothly pass through this
singularity without any catastrophic implications on the physical quantities.
As was shown in \cite{Odintsov:2015gba}, the graceful exit from the
inflationary era may be triggered by this type of soft singularity.
\end{itemize}
Here, $\rho_\mathrm{eff}$ and $p_\mathrm{eff}$ are defined by
\begin{equation}
\label{IV}
\rho_\mathrm{eff} \equiv \frac{3}{\kappa^2} H^2 \, , \quad
p_\mathrm{eff} \equiv - \frac{1}{\kappa^2} \left( 2\dot H + 3 H^2
\right)\, .
\end{equation}
Note that $\rho_\mathrm{eff}$ and $p_\mathrm{eff}$ are different from
$\rho$ and $p$ in (\ref{JGRG11}).
For example, $\rho_\mathrm{eff}$ and $p_\mathrm{eff}$ may
include the contribution from the modified gravity.
Then Eq.~(\ref{IV}) shows that for Type I and III singularities, $H$ diverges
but for Type II and IV, $H$ is finite.
However, in Type II singularity  $\dot H$ diverges.
We will be interesting in the account of thermal effects especially for
Type I and III singularities.

There is also a scenario called the Little Rip cosmology
\cite{Frampton:2011sp,Brevik:2011mm,Frampton:2011rh} where the universe enters
to singular state at infinite future.
In this scenario, the Hubble rate is finite in the finite time but it becomes
infinite in the infinite future.
We show that thermal radiation will become important in the far future for
the Little Rip evolution much before the arrival to the infinite Rip time.

\subsection{Big Rip with thermal effects: transition to Type II singularity}

The Big Rip singularity of the universe can be generated by the cosmic fluid,
which is often called ``phantom'',  with the equation of state (EoS)
parameter $w$, which is defined by
\begin{equation}
\label{EoS}
w= \frac{p}{\rho} \, ,
\end{equation}
for the pressure $p$ and the energy density $\rho$ for general cosmic fluid, 
it is less than $-1$, $w<-1$.
By assuming the conservation law,
\begin{equation}
\label{consv}
0=\dot\rho + 3 H \left( \rho + p \right)\, ,
\end{equation}
we find
\begin{equation}
\label{density}
\rho = \rho_0 a^{-3 \left( 1 + w \right)} \, ,
\end{equation}
with a positive constant $\rho_0$.
Then in case of the phantom, because $-3 \left( 1 + w \right)>0$, the energy
density dominates at late time where $a$ becomes large.
Then by using the FRW equations (\ref{JGRG11}), we find $H$ behaves as
\begin{equation}
\label{rip}
H \propto \frac{1}{t_\mathrm{Rip} - t} \, ,
\end{equation}
and $H$ diverges at $t=t_\mathrm{Rip}$, which is the Big Rip singularity.

Near the Big Rip singularity, the temperature of the universe becomes large and
we may expect the generation of the thermal radiation as in the case 
of the Hawking radiation.
The Hawking temperature $T$ is proportional to the inverse of the radius
$r_\mathrm{H}$ of the apparent horizon and the radius $r_\mathrm{H}$ 
is proportional to the inverse of the Hubble rate $H$.
Therefore the temperature $T$ is proportional to the Hubble rate $H$.
As well-known in the statistical physics, the energy-density 
$\rho_\mathrm{t\_rad}$ of the thermal radiation is 
proportional to the fourth power of the temperature.
Then when $H$ is large enough, we may assume that the energy-density of the
thermal radiation is given by
\begin{equation}
\label{BRHR1}
\rho_\mathrm{t\_ rad} = \alpha H^4 \, ,
\end{equation}
with a positive constant $\alpha$.
At the late time, the FRW equation (\ref{JGRG11}) should be modified 
by the account of thermal radiation,
\begin{equation}
\label{BRHR2}
\frac{3}{\kappa^2} H^2 = \rho_0 a^{-3\left( 1 + w \right)} + \alpha H^4 \, .
\end{equation}
Here we assume $w<-1$.
At the late time but much before the Big Rip time, the first term in the
equation (\ref{BRHR2}) dominates and therefore the universe expands to 
the Big Rip singularity, where the Hubble rate $H$
behaves as in (\ref{rip}).
Then near the Big Rip time $t_\mathrm{Rip}$, the second term in (\ref{BRHR2})
should dominate and we obtain
\begin{equation}
\label{BRHR3}
\frac{3}{\kappa^2} H^2 \sim \alpha H^4 \, ,
\end{equation}
whose non-trivial solution is given by
\begin{equation}
\label{BRHR4}
H^2 = H_\mathrm{crit}^2 \equiv \frac{3}{\kappa^2 \alpha} \, .
\end{equation}
As $H$ goes to a constant, we might expect that the space-time goes to
the asymptotically de Sitter space-time but it is not true.
Even in the de Sitter space-time, the scale factor $a$ becomes larger and
larger as an exponential function of $t$, then the first term in the equation 
(\ref{BRHR2}) should dominate finally.
The Hubble rate $H$ is, however, already larger than $H_\mathrm{crit}$, 
there is no solution of  (\ref{BRHR2}).
Then the universe should end up at finite time with some kind of the
singularity.

For more quantitative analysis, we solve (\ref{BRHR2}), with respect to $H^2$
as follows,
\begin{equation}
\label{BRHR5}
H^2 = \frac{\frac{3}{\kappa^2} \pm \sqrt{\frac{9}{\kappa^4} - 4\alpha \rho_0
a^{-3\left( 1 + w \right)}}}
{2\alpha} \, .
\end{equation}
Because $H^2$ is a real number, we find that there is a maximum for the scale
factor $a$,
\begin{equation}
\label{BRHR6}
a \leq a_\mathrm{max} \equiv \left( \frac{9}{4 \kappa^4 \alpha \rho_0}
\right)^{- \frac{1}{3\left( 1 + w \right)}} \, .
\end{equation}
Then we consider the behavior of $a$ or $H$ around the maximal
$a=a_\mathrm{max}$ by writing the scale factor $a$ as
\begin{equation}
\label{BRHR7}
a = a_\mathrm{max} \e^N \, .
\end{equation}
Here $N$ corresponds to the $e$-folding number but $N$ should be negative
because $a<a_\mathrm{max}$.
Furthermore because we are interested in the region $a\sim a_\mathrm{max}$, we
assume $\left| N \right| \ll 1$.
Then by using $H=\frac{dN}{dt}$,  Eq.~(\ref{BRHR5}) can be rewritten as
\begin{equation}
\label{BRHR8}
\left( 1 \mp \frac{1}{2} \sqrt{3\left( 1 + w \right)N} \right) dN \sim dt
\sqrt{ \frac{3}{2\alpha \kappa^2}} \, ,
\end{equation}
which can be integrated as
\begin{equation}
\label{BRHR9}
N \mp \frac{1}{3} \left( -N\right)^\frac{3}{2} \sqrt{-3\left( 1 + w
\right)} \sim - \left( t_\mathrm{max} - t \right) \sqrt{ \frac{3}{2\alpha
\kappa^2}} \, .
\end{equation}
Here $a=a_\mathrm{max}$ when $t=t_\mathrm{max}$.
Because we are assuming $\left| N \right| \ll 1$, Eq.~(\ref{BRHR9}) can be
rewritten as
\begin{equation}
\label{BRHR10}
N \sim - \left( t_\mathrm{max} - t \right) \sqrt{ \frac{3}{2\alpha \kappa^2}}
\mp \frac{\sqrt{-3\left( 1 + w \right)}}{3} \left( \left( t_\mathrm{max} - t
\right) \sqrt{ \frac{3}{2\alpha \kappa^2}} \right)^\frac{3}{2} \, .
\end{equation}
Because $H=\frac{dN}{dt}$, we find
\begin{align}
\label{BRHR11}
H \sim& \sqrt{ \frac{3}{2\alpha \kappa^2}}
\mp \frac{\sqrt{-3\left( 1 + w \right)}}{2} \left(
\sqrt{ \frac{3}{2\alpha \kappa^2}} \right)^\frac{3}{2}
\left( t_\mathrm{max} - t \right)^{\frac{1}{2}} \, , \nn
\dot H \sim& \mp \frac{\sqrt{-3\left( 1 + w \right)}}{4} \left(
\sqrt{ \frac{3}{2\alpha \kappa^2}} \right)^\frac{3}{2}
\left( t_\mathrm{max} - t \right)^{-\frac{1}{2}} \, .
\end{align}
Then in the limit $t\to t_\mathrm{max}$, although $H$ is finite but
$\dot H$ diverges.
Therefore the universe ends up with Type II singularity at $t=t_\mathrm{max}$
and the cyclic cosmology does not occur. Thus, we demonstrated that the account of
thermal effects near the Big Rip singularity changes the universe evolution to
the finite-time Type II singularity.

\subsection{Type III singularity with account of thermal effects: transition to
Type II singularity}

The scale factor which generates Type III singularity
can be expressed as
\begin{equation}
\label{III1}
a(t) = a_s \e^{ - \beta \left( t_s - t \right)^\gamma} \, ,
\end{equation}
with positive constants $a_s$, $t_s$, $\beta$, and $\gamma$.
In order to generate Type III singularity we restrict the value
of $\gamma$ as
\begin{equation}
\label{III2}
0<\gamma<1 \, .
\end{equation}
Then the Hubble rate $H$ is given by
\begin{equation}
\label{III3}
H = \beta \gamma \left( t_s - t \right)^{\gamma -1} \, .
\end{equation}
Hence, in the limit $t\to t_s$, $H$ diverges but the scale factor $a$
is finite. 
 From Eq.~(\ref{IV}) it follows
\begin{equation}
\label{III4}
\rho_\mathrm{eff} = \frac{3}{\kappa^2}
\beta^2 \gamma^2 \left( t_s - t \right)^{2 \left(\gamma -1\right)}
\, , \quad
p_\mathrm{eff} = - \frac{1}{\kappa^2}
\left( - 2 \beta \gamma \left( \gamma - 1 \right)
\left( t_s - t \right)^{\gamma -2}
+ 3 \beta^2 \gamma^2 \left( t_s - t
\right)^{2 \left(\gamma -1\right)}
\right)\, .
\end{equation}
By deleting $\left( t_s - t \right)$, we find the following
equation of state,
\begin{equation}
\label{III5}
p_\mathrm{eff} = - \rho_\mathrm{eff}
+ \frac{2 \beta \gamma \left( \gamma - 1 \right)}{\kappa^2}
\left( \frac{\kappa^2 \rho_\mathrm{eff}}{3 \beta^2 \gamma^2}
\right)^{\frac{\gamma - 2}{2 \left(\gamma -1\right)}}\, .
\end{equation}
Using the conservation law (\ref{consv}) or directly
using (\ref{III1}) and (\ref{III4}), one gets
\begin{equation}
\label{III6}
\rho_\mathrm{eff} = \frac{3}{\kappa^2}
\beta^2 \gamma^2 \left( \frac{1}{\beta} \ln \left(
\frac{a_s}{a(t)}\right)
\right)^{\frac{2 \left(\gamma -1\right)}{\gamma}}
= \frac{3\gamma^2 \beta^{\frac{2}{\gamma}}}{\kappa^2}
\left( \ln \left( \frac{a_s}{a(t)}\right)
\right)^{\frac{2 \left(\gamma -1\right)}{\gamma}}
\, .
\end{equation}
With the account of the thermal radiation, instead of
(\ref{BRHR2}), we have
\begin{equation}
\label{III7}
\frac{3}{\kappa^2} H^2 = A \left( \ln \left( \frac{a_s}{a(t)}\right)
\right)^{-B} + \alpha H^4 \, , \quad
A \equiv \frac{3\gamma^2 \beta^{\frac{2}{\gamma}}}{\kappa^2} \, ,
\quad B \equiv - \frac{2 \left(\gamma -1\right)}{\gamma} > 0 \, .
\end{equation}
Then instead of (\ref{BRHR5}), we obtain
\begin{equation}
\label{III8}
H^2 = \frac{\frac{3}{\kappa^2} \pm \sqrt{\frac{9}{\kappa^4} - 4\alpha
A \left( \ln \left( \frac{a_s}{a(t)}\right) \right)^{-B}}}
{2\alpha} \, .
\end{equation}
Then in order that $H^2$ to be real, we find that there is
a maximum $a_\mathrm{max}$ for $a(t)$,
\begin{equation}
\label{III9}
a(t) \leq a_\mathrm{max} \equiv a_s
\e^{-\left(\frac{9}{4A \alpha \kappa^2}\right)^{- \frac{1}{B}}}
< a_s \, .
\end{equation}
Because $a_\mathrm{max}$ is smaller than $a_s$, we find that dark universe with
the future Type III singularity is transited to the one with Type II
singularity due to the account of thermal effects.

\subsection{Thermal radiation for Type II and Type IV singularities}

In case of Type II and Type IV singular universes,
the Hubble rate $H$  behaves as
\begin{equation}
\label{R13B}
H \sim H_0 + h_0 \left(t_s - t\right)^{-\beta}\, .
\end{equation}
When $ 0> \beta > -1$ the behavior of $H$ corresponds to Type II
and when $\beta < -1$ but $\beta$ is not an integer, to Type IV.

When one considers  general matter,
the first FRW equation where
usual matter and the thermal radiation as in (\ref{BRHR2}) are included,
is given by
\begin{equation}
\label{CVT01B}
\frac{3}{\kappa^2} H^2 = \rho + \alpha H^4 \, .
\end{equation}
Here $\rho$ is matter energy-density.
In case of Type II or Type IV singularity, if $H_0\neq 0$,
near the singularity, the l.h.s. goes to a finite value
$\frac{3}{\kappa^2} H^2 \to \frac{3}{\kappa^2} H_0^2$
and the contribution from the thermal radiation in the r.h.s. also
becomes finite, $\alpha H^4 \to \alpha H_0^4$.
Therefore the thermal radiation does not change the structure of the
singularity.
If $H_0=0$, the r.h.s. behaves as $\left(t_s - t\right)^{-4\beta}$ and the
contribution from the thermal radiation behaves as 
$\left(t_s - t\right)^{-2\beta}$.
Because $\beta<0$, the contribution from the
thermal radiation is less dominant and therefore the thermal radiation does not
change the structure of the singularity.

\subsection{Little Rip universe with the account of thermal effects}

In the qualitatively-different from the above ones, the Little Rip scenario
\cite{Frampton:2011sp,Brevik:2011mm,Frampton:2011rh},
the Hubble rate $H$ becomes infinite at the infinite future.
A simple example is given by
\begin{equation}
\label{LR1}
H = H_0 t \, ,
\end{equation}
with positive $H_0$.
Then 
\begin{equation}
\label{LR2}
\rho_\mathrm{eff} = \frac{3}{\kappa^2} H_0^2 t^2 \, , \quad
p_\mathrm{eff} = - \frac{1}{\kappa^2} \left( 2 H_0 + 3 H_0^2 t^2
\right)\, .
\end{equation}
The equation of state is given by
\begin{equation}
\label{LR3}
p_\mathrm{eff} = - \rho_\mathrm{eff} - \frac{2 H_0}{\kappa^2} , .
\end{equation}
Eq.~(\ref{LR1}) also shows that the scale factor $a(t)$ is given by
\begin{equation}
\label{LR4}
a(t) = a_0 \e^{\frac{1}{2} H_0 t^2} \, ,
\end{equation}
with a constant $a_0$.
Then Eq.~(\ref{LR2}) shows
\begin{equation}
\label{LR5}
\rho_\mathrm{eff} = \frac{6H_0}{\kappa^2}\ln \left( \frac{a(t)}{a_0} \right) 
\, .
\end{equation}
When we include the contribution from the thermal radiation,
the equation corresponding to (\ref{BRHR2}) or (\ref{III7}) has the following
form,
\begin{equation}
\label{LR6}
\frac{3}{\kappa^2} H^2 = \frac{6H_0}{\kappa^2}\ln \left( \frac{a(t)}{a_0}
\right)
+ \alpha H^4 \, ,
\end{equation}
and the equation corresponding to (\ref{BRHR5}) or (\ref{III8}) has the
following form,
\begin{equation}
\label{LR7}
H^2 = \frac{3}{2\alpha\kappa^2} \left( 1 \pm \sqrt{1
 - \frac{2\alpha H_0\kappa^2}{3} \ln \left( \frac{a(t)}{a_0} \right)} \right)
\, .
\end{equation}
Again, from Eq.~(\ref{LR7}) it follows  that there is a maximum of the scale
factor $a$,
\begin{equation}
\label{LR8}
a(t) \leq a_\mathrm{max} \equiv a_0 \e^{\frac{3}{2\alpha\kappa^2}} \, ,
\end{equation}
and therefore the space-time corresponds to Type II singularity.
Thus we again see that due to the account of thermal effects, the dark energy
which should bring the universe to the Little Rip at the infinite future
changes its evolution to Type II singularity. The corresponding
transition occurs!

\section{Future singularities with account of quantum and thermal effects}

In the previous sections, we have shown that any scenario, where the Hubble
rate becomes infinite in the finite or infinite future as for Type I (Big
Rip) and Type III singularities and in the Little Rip universe scenario, 
will not be realized if we include the effect from the thermal radiation. 
The universe will change its evolution to Type II singularity.
From other side when the universe approaches to the future singularity, its
curvature and other geometrical invariants grow up. 
As a result, the quantum effects may change the behavior of the future 
space-time singularity.
For example, one can show that quantum effects may change the structure of
future singularity, see \cite{Nojiri:2004ip,Nojiri:2004pf}
(see also \cite{Kamenshchik:2013naa,Bates:2010nv,Tretyakov:2005en,
Calderon:2004bi,Carlson:2016iuw}).
In this section, we use simple qualitative arguments of
Ref.~\cite{Nojiri:2010wj} to show the role of quantum effects in
conformally-invariant theories to future singularity and compare it with the
effect due to thermal radiation.

As is well-known, the conformal anomaly $T_A$ has the following form:
\begin{equation}
\label{OVII}
T_A=b\left(\mathcal{F} + \frac{2}{3}\Box R\right) + b' \mathcal{G}
+ b''\Box R\, .
\end{equation}
Here $\mathcal{F}$ is the square of the 4D Weyl tensor, and
$\mathcal{G}$ is the Gauss-Bonnet invariant, which are given by
\begin{equation}
\label{GF}
\mathcal{F} = \frac{1}{3}R^2 -2 R_{\mu\nu}R^{\mu\nu}
+ R_{\mu\nu\rho\sigma}R^{\mu\nu\rho\sigma}\, , \quad
\mathcal{G}=R^2 -4 R_{\mu\nu}R^{\mu\nu}
+ R_{\mu\nu\rho\sigma}R^{\mu\nu\rho\sigma}\, .
\end{equation}
In case that matter is conformally-invariant and there appear $N$ scalars,
$N_{1/2}$ spinors, $N_1$ vector fields, $N_2$
($=0$ or $1$) gravitons, and $N_\mathrm{HD}$ higher-derivative conformal
scalars, $b$ and $b'$ have the following forms,
\begin{equation}
\label{bs}
b= \frac{N +6N_{1/2}+12N_1 + 611 N_2 - 8N_\mathrm{HD}}{120(4\pi)^2}
\, ,\quad
b'=- \frac{N+11N_{1/2}+62N_1 + 1411 N_2 -28 N_\mathrm{HD}}{360(4\pi)^2}\ .
\end{equation}
As is shown in (\ref{bs}), $b$ is positive and $b'$ is negative for the usual
matter.
An exception is the higher-derivative conformal scalar.
The value of $b''$ can be always shifted by adding $R^2$ to the classical
action.

If we write the energy density $\rho_A$ and pressure $p_A$
corresponding to the trace anomaly $T_A$, we find
$T_A=- \rho_A + 3p_A$.
Then by using the energy conservation law in the FRW universe
\begin{equation}
\label{CB1}
0=\frac{d\rho_A}{dt} + 3 H\left(\rho_A + p_A\right)\, ,
\end{equation}
we can delete $p_A$ as
\begin{equation}
\label{CB2}
T_A=-4\rho_A - \frac{1}{H}\frac{d\rho_A}{dt}\, ,
\end{equation}
which can be integrated and we find the following expression for $\rho_A$
\cite{Nojiri:2005sx}:
\begin{equation}
\label{CB3}
\rho_A = -\frac{1}{a^4} \int dt a^4 H T_A \, .
\end{equation}
By using the above expression and identifying $\rho_\mathrm{eff}=\rho_A$,
one may consider the FRW equation~(\ref{IV}).

However, as in \cite{Nojiri:2010wj}, for simplicity,
we consider the trace of the Einstein equation
including the trace anomaly, as follows,
\begin{equation}
\label{CA1}
R = - \frac{\kappa^2}{2} \left(T_\mathrm{matter} + T_A \right)\, .
\end{equation}
Here $T_\mathrm{matter}$ is the trace of the matter energy-momentum tensor.
For the FRW universe (\ref{JGRG14}), $\mathcal{F}$ and $\mathcal{G}$ are given
by
\begin{equation}
\label{CA2}
\mathcal{F}=0\ ,\quad \mathcal{G}=24\left(\dot H H^2 + H^4\right)\ .
\end{equation}
{
What we like to show is that if there is a singularity, the trace equation (\ref{CA2})
cannot be consistent.
Especailly we show that the contribution from the conformal anomaly in the
r.h.s. of Eq.~(\ref{CA1}) is more singular than the scalar curvature in the l.h.s.
If the contribution from the matter although the conformal anomaly may give some
corrections. Note that rigorous study of the account of quantum effects may
be done following Ref.~\cite{Carlson:2016iuw} but it requests numerical study depending of
particles content of the universe as well as effective dark fluid.
}

We now assume that $H$ behaves as
\begin{equation}
\label{R13}
H \sim H_0 + h_0 \left(t_s - t\right)^{-\beta}\, .
\end{equation}
When $\beta\geq 1$, the behavior of $H$ corresponds to Type I
(Big Rip) singularity, and when $1\geq \beta > 0$, to Type III,
when $0> \beta > -1$ to Type II, when $\beta < -1$ but $\beta$ is not an
integer, to Type IV singularity.
One may also neglect the contribution from matter and put
$T_\mathrm{matter}=0$.

In case that $\beta>0$, which corresponds to Type I (Big Rip) and 
Type III singularity,
the first constant term $H_0$ in (\ref{R13}) seems to be less dominant
and we may neglect this term.
Then because the scalar curvature $R$ is given by $R=12H^2 + 6\dot H$,
when $\beta \geq 1$ (Type I), we find that the scalar curvature $R$ behaves as
$R\sim \left(t_s - t\right)^{-2\beta}$ and when $0<\beta<1$ (Type III), $R$
behaves as $R\sim \left(t_s - t\right)^{-\beta-1}$.
On the other hand, when $-1<\beta<0$ (Type II), $R$ behaves as
$R\sim \left(t_s - t\right)^{-\beta-1}$.
When $\beta<-1$, which corresponds to Type IV singularity if $\beta$ 
is not an integer, if $H_0\neq 0$, $R$ behaves as a constant but if $H_0=0$,
$R\sim \left(t_s - t\right)^{-\beta-1}$.

We now assume the behavior of the Hubble rate as in (\ref{R13}).
Then in case of Type I (Big Rip) case,
where $\beta\geq 1$, near the Big Rip singularity,
$t\sim t_s$, as seen from (\ref{CA2}), the Gauss-Bonnet invariant
$\mathcal{G}$ behaves as
$\mathcal{G} \sim 24 H^4 \sim \left(t_s - t\right)^{-4\beta}$
and therefore $\mathcal{G}$ becomes very large and the contribution
from the matter $T_\mathrm{matter}$ in (\ref{CA1})
can be neglected.
On the other hand, one finds $\Box R \sim \left(t_s - t\right)^{-2\beta -2}$.
Then because $R\sim \left(t_s - t\right)^{-2\beta}$, $T_A$ becomes much larger
than $R$ and therefore Eq.~(\ref{CA1}) cannot be satisfied.
This shows that the quantum effects coming from the conformal anomaly also
remove Type I (Big Rip) singularity.

In case of Type II singularity, where $-1<\beta<0$,
we find that $\mathcal{G}$ behaves as
$\mathcal{G}\sim 24 \dot H H^2 \sim \left(t_s - t \right)^{-3\beta -1}$.
Because $R \sim \left(t_s - t \right)^{-\beta - 1}$, 
the Gauss-Bonnet term in $T_A$ is less singular and therefore negligible
compared with $R$ and the contribution from the matter.
Therefore, the Gauss-Bonnet term in $T_A$ does not help to prevent 
Type II singularity.
Note, however, $\Box R$ behaves as
$\Box R \sim \left(t_s - t\right)^{-\beta - 3}$, which is more singular than
the scalar curavature. Then if $2b/3 + b''\neq 0$,
the contribution from $T_A$ becomes mich larger than $R$ near the
singularity $t\sim t_s$ and Eq.~(\ref{CA1}) cannot be satisfied.
Therefore if $2b/3 + b''\neq 0$, even Type II singularity can be also
prevented when  the quantum effects due conformal
anomaly are included.

In case of Type III singularity ($0<\beta<1$), the Gauss-Bonnet invariant
behaves as
$\mathcal{G}\sim 24 \dot H H^2 \sim \left(t_s - t \right)^{-3\beta -1}$
and $\Box R$ behaves as $\Box R \sim \left(t_s - t\right)^{-\beta - 3}$.
Because the scalar curvature behaves as
$R \sim \left(t_s - t \right)^{-\beta - 1}$, both of the terms, $\Box R$ and
$\mathcal{G}$, are more singular than the scalar curvature $R$ and Type III singularity
is also prevented. Thus, we demonstrated that quantum effects may remove
finite-time future singularities. Note that account of quantum gravity effects
in specific models also is known to remove the Big Rip singularity
\cite{Elizalde:2004mq}.

Let us include the thermal effects to above analysis.
So far  the trace part of the Einstein equation is used.
As the radiation is usually conformal, the trace part of the energy-momentum
tensor of the radiation should vanish and the thermal radiation does not
contribute to the trace equation.
We should be, however, more careful in the present situation.
The energy-density of the thermal radiation is only determined by the
temperature.
Therefore, the universe expands and its volume with the thermal radiation
increases, the total energy should also be increased if the temperature is
not changed or increases as in the case of Type I (Big Rip) or Type III singularity,
or the Little Rip cosmology.
In other words, say,  in the phantom universe, there should exist effectively
negative pressure.
The energy of the thermal radiation is not conserved because the
expansion produces the new thermal radiation.
We should note, however, in order that the effective pressure, which includes
the effect of the expansion, is consistent with the FRW equations,
the energy-density of the thermal radiation and the effective pressure must
satisfy the conservation law
\begin{equation}
\label{CVT1}
0=\frac{d\rho_\mathrm{t\_rad}}{dt} + 3 H\left(\rho_\mathrm{t\_rad}
+ p_\mathrm{t\_rad}\right)\, .
\end{equation}
To show the conservation law, we may start from the first FRW equation where
usual matter and the thermal radiation as in (\ref{BRHR2}) are included,
\begin{equation}
\label{CVT01}
\frac{3}{\kappa^2} H^2 = \rho + \rho_\mathrm{t\_rad} \, , \quad
\rho_\mathrm{t\_rad} = \alpha H^4 \, .
\end{equation}
Here $\rho$ is matter energy-density.
By considering the derivative of Eq.~(\ref{CVT01}) with respect to time $t$,
we obtain,
\begin{equation}
\label{CVT02}
\frac{6}{\kappa^2} H \dot H = \dot\rho + 4 \alpha H^3 \dot H\, .
\end{equation}
Then by using the standard conservation law for  matter,
\begin{equation}
\label{CVT03}
0=\dot\rho + 3 H \left( \rho + p \right) \, ,
\end{equation}
with the matter pressure, and combining (\ref{CVT01}) and
(\ref{CVT02}), we obtain
\begin{equation}
\label{CVT04}
 - \frac{1}{\kappa^2} \left( 2\dot H + 3 H^2 \right) = p
 - \alpha \left( H^4 + \frac{4}{3}H^2 \dot H \right) \, ,
\end{equation}
which is nothing but the second FRW equation and we can identify
the effective pressure of the thermal radiation as follows,
\begin{equation}
\label{CVT2}
p_\mathrm{t\_rad} = - \alpha \left( H^4 + \frac{4}{3}H^2 \dot H \right)\, .
\end{equation}
Thus effectively, the energy-density and the
effective pressure of the thermal radiation satisfy the conservation law
(\ref{CVT1}) or we can find the exact and unique form of the effective pressure
in (\ref{CVT2}) directly by using the conservation law (\ref{CVT1}) and assuming
the form of the energy-density of the radiation in (\ref{BRHR1}).

Then the trace part
$T_\mathrm{t\_rad} = - \rho_\mathrm{t\_rad} + 3 p_\mathrm{t\_rad}$
of the energy-momentum tensor for the radiation including the effect of the
expansion of the universe is given by
\begin{equation}
\label{CVT2BB}
T_\mathrm{t\_rad} = -4 \alpha \left( H^4 + H^2 \dot H \right) \, .
\end{equation}
Let us  assume the behavior of the Hubble rate $H$ as in (\ref{R13}).
Then in case of Type I (Big Rip) case ($\beta\geq 1$), near the singularity,
we find $T_\mathrm{t\_rad} \sim \left(t_s - t\right)^{-4\beta}$,
whose behavior is not so changed from that of $T_A$ although we need to
compare $b'$ with $\alpha$ to see which term is dominant one.

In case of Type II singularity ($-1<\beta<0$),
we find $T_\mathrm{t\_rad} \sim \left(t_s - t \right)^{-3\beta -1}$.
As $R \sim \left(t_s - t \right)^{-\beta - 1}$,
the contribution  from $T_\mathrm{t\_rad}$ is negligible.
In case $b''\neq 0$, which is arbitary and can be put to vanish if we dont add
$R^2$ term, the contribution from $\Box R$ in $T_A$ dominates and the 
Type II singularity does not occur.

In case of Type III singularity ($0<\beta<1$), 
we find $T_\mathrm{t\_rad} \sim \left(t_s - t \right)^{-3\beta -1}$,
whose behavior is not changed from that of the Gauss-Bonnet invariant
$\mathcal{G}$ in $T_A$ but weaker than the behavior of $\Box R$.
Then if $b''\neq 0$, the contribution from the thermal radiation is
less dominant than that of the conformal anomaly $T_A$.
If $b''=0$, the contribution from $T_\mathrm{t\_rad}$ is not changed from that
from $T_A$ and we need to compare $b'$ with $\alpha$ to see which could be
dominant, again.
Thus, we demonstrated that when quantum effects dominate over thermal effects
then future singularities are removed. However, in some cases which depend on
the specific features of the theory under consideration the dominant contribution
is due to thermal effects. In this case,
the most possible universe future is Type II singularity.

\section{Summary}

In summary, we studied the singular dark universe future where singularity is
caused by the corresponding dark fluid while also thermal effects due to
the Hawking radiation on apparent horizon of the FRW universe are included.
It is shown that for dark universe with Type I, III singularities
and for the Little Rip universe the transition to Type II singularity at final
state occurs. On the same time for Type II and IV singular universe one sees no
qualitative effect due to thermal radiation. When in addition to thermal
radiation also quantum effects are taken into account the situation is more
complicated. Usually, matter quantum effects (at least, for
conformally-invariant fields) remove the finite-time future singularity.
Together with thermal effects the universe future is defined by  which
of terms (thermal or quantum) in the effective energy-density is dominant. This
depends from the specific features of the universe under consideration (fields
content, fluids, coefficient of thermal energy-density, etc.). In particulary,
when thermal effects are dominant, then the future universe state is Type II
singularity, again. 

Some remark is in order. It is known that several million years before the Rip time, 
there appears some inertial force which may unbound particles producing
desintegration of all bound objects at the universe. Let us check the effect of
thermal radiation to this inertial force.
A test mass $m$, which is separated from an observer by the distance $r$,
receives an inertial force of tidal force $F_\mathrm{in}$ when the observer
observes the mass, as follows,
\begin{equation}
\label{TuA07}
F_\mathrm{in} = r m \frac{\ddot a}{a} = r m \left( \dot H+ H^2 \right) \, .
\end{equation}
In case of Type I (Big Rip) or Type III singularity, because $H$ and $\cot H$
become very large near the singulrity and therefore any
extended object will be ripped and destroyed.
If we take into account the contribution from the thermal radiation, the
singularity will reduces to Type II singularity, where although $H$ is finite or vanish,
$\dot H$ and therefore the inertial force becomes very large near the singularity.
Hence, in this case the bound objects are desintegrated as in the case without
thermal effects.
In case of Type IV singularity, both of $H$ and $H'$ are finite and therefore
the inertial force is also finite.
Finally, it may be of interest to study the role of thermal radiation for future singularities
in modified gravity theories. This will be done elsewhere.

\begin{acknowledgments}
This work is partially supported  by MEXT KAKENHI Grant-in-Aid for
Scientific Research on Innovative Areas ``Cosmic Acceleration'' No. 15H05890
(S.N.) and the JSPS Grant-in-Aid for Scientific Research (C) No. 18K03615
(S.N.).This work was done partially in the framework of the Russian
Government Program of Competitive Growth of the Kazan Federal University.
\end{acknowledgments}

\end{document}